\documentstyle[11pt,newpasp,twoside,epsf]{article}

\def\edcomment#1{\iffalse\marginpar{\raggedright\sl#1\/}\else\relax\fi}
\reversemarginpar

\begin{document}
\title{A search for disks around massive young stellar objects}
 \author{A.G. Gibb}
\affil{Univ of Maryland, Dept of Astronomy, College Park, MD 20742, USA}
\author{M.G. Hoare}
\affil{Univ of Leeds, Dept of Physics and Astronomy, Leeds, LS2 9JT, UK}
\author{L.G. Mundy}
\affil{Univ of Maryland, Dept of Astronomy, College Park, MD 20742, USA}
\author{F. Wyrowski}
\affil{Max Planck Institut f\"ur Radioastronomie, Auf dem H\"ugel 69,
 Bonn D-53121, Germany}

\begin{abstract}
We present subarcsecond observations at 2.7 and 1.4 mm of a sample of
massive young stellar objects made with the BIMA millimetre array.
For most sources the continuum emission on the smallest scales at 2.7
mm is dominated by free-free emission from the stellar wind or jet.
Strong emission at 1.4 mm shows the presence of significant dust
associated with Cep A and GL\,490 but our resolution is not sufficient
to resolve any structure. The 2.7-mm emission from GL\,490 is resolved
but it is not clear whether we are seeing a single circumstellar disk
or a secondary companion, although near-infrared data support the disk
hypothesis. Estimates of the dust mass yield values of $\sim$1--4
M$_\odot$ within radii of 150 to 1000 AU.
\end{abstract}

\section{Introduction}

The existence of circumstellar disks around low-mass protostars is now
well-established (Mundy, Looney \& Welch 2000). However, the situation
for their massive counterparts remains unclear. Indirect observational
evidence for circumstellar disks around massive young stellar objects
(YSOs) comes from the presence of bipolar outflows, polarization
patterns and the presence of ionized winds extending in an equatorial
direction (Hoare 2002). In one case, a model of the 7-mm emission
shows that the data are consistent with a circumstellar disk with a
diameter of 130 AU (Shepherd, Claussen \& Kurtz 2001). Molecular
emission line signatures of rotation have mostly been confined to
larger and therefore non-circumstellar scales (e.g.\ Zhang, Hunter, \&
Sridharan 1998).

The question of whether or not circumstellar disks are present around
massive stars is key in current theoretical debate. One important
question is whether high-mass star formation is simply a scaled-up
version of low-mass star formation where accretion onto the central
star occurs via a disk. Direct observation of disks around massive
YSOs is key to testing the theories for how massive stars form. It is
therefore important to determine the circumstellar mass distribution
around luminous YSOs.

\section{Observations}

We have used the Berkeley-Illinois-Maryland Association (BIMA: see
Welch et al.\ 1996) millimetre array to observe a sample of 8 massive
young stellar objects: GL\,490, Cep A, NGC7538-IRS1, NGC7538-IRS9,
G35.2$-$0.7N, GGD27-IRS, W75N and GL\,2591. The observations were made
in continuum bands at 2.7 mm (107 GHz) for all sources (except
GGD27-IRS which was observed at 95 GHz) and 1.4 mm (216 GHz) for
GL\,490, Cep A and NGC7538-IRS1 only. Data were conventionally
calibrated using a nearby phase calibrator, while fast-switching was
employed in the more extended configurations (see Looney, Mundy \&
Welch 2000). In three sources (Cep A, NGC7538-IRS1 and G35.2$-$0.7N),
the 107-GHz methanol maser was sufficiently strong to apply a
self-calibration solution (the 95-GHz methanol maser was used for
GGD27-IRS).  The configurations used were A and B (at 2.7 mm) and B
and C (at 1.4 mm), yielding an angular resolution varying from 0.3
arcsec (for Cep A, GL\, 490 and NGC7538-IRS1) to approximately 1
arcsec at 2.7 mm, which translates to a radius of 105 to 350 AU at the
distance to our nearest source (Cep A). At 1.4 mm the angular
resolution was $\sim$1.5--2.0 arcsec. The typical sensitivity was 3--8
mJy\,bm$^{-1}$ at 2.7 mm, rising to 20--30 mJy\,bm$^{-1}$ at 1.4 mm.

\section{Results}

\subsection{GL\,490}

Figure 
1 
shows the 2.7-mm and 1.4-mm continuum emission towards
GL\,490. The 2.7-mm emission arises from an elongated structure with
dimensions 0.7$\times$0.5 arcsec$^2$, corresponding to 700$\times$500
AU$^2$ at a distance of 1 kpc. The emission peaks on the position of
the VLA source with an extension 0.3-arcsec to the south-east. It is
not clear whether this represents part of a circumstellar disk around
GL\,490 or if it marks the location of a secondary companion. However,
we note that the high-resolution $H$-band speckle-interferometry image
of Hoare, Glindemann \& Richichi (1996) shows no evidence for
multiplicity reinforcing the disk interpretation. The position angle
of the 2.7-mm emission is in good agreement with the larger structures
observed by Mundy \& Adelman (1988) and more recently by Schreyer et
al.\ (2002).

An image made using only the longest baselines yields a weak point
source centred on the VLA position. The flux of this source is in good
agreement with the predicted flux of the wind contribution at 2.7 mm
indicating that the flux is dominated by the ionized wind on scales of
0.1 arcsec (100 AU) or less.

On the other hand, the 1.4-mm emission is dominated by dust, although
our resolution is not high enough to resolve any structure. The image
shown in Fig. 
1b
is consistent with a point source within
the 2.4$\times$1.1 arcsec$^2$ beam. 

\begin{figure}
\plotone{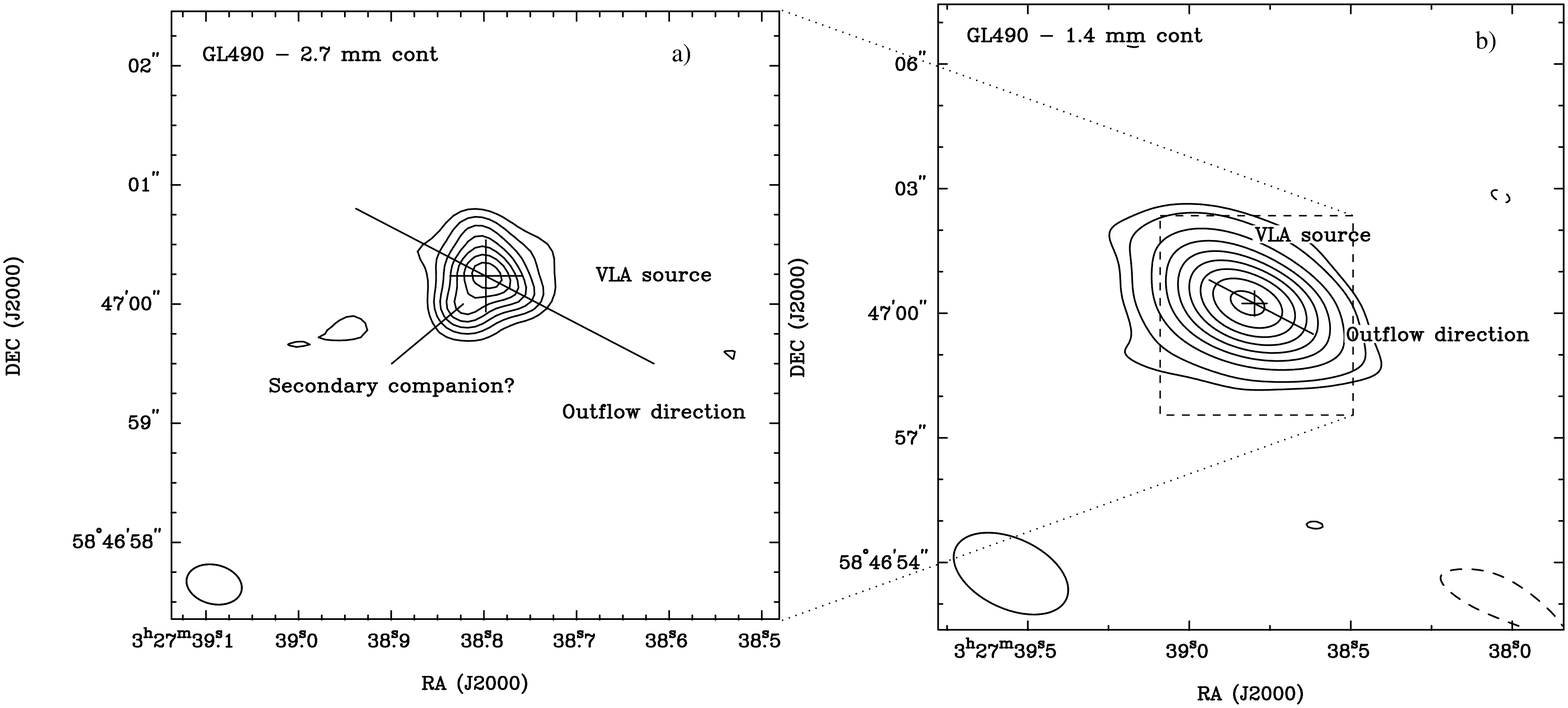}
\caption{Continuum emission towards GL\,490 at a) 2.7-mm and b) 1.4
  mm. The radio source is marked by a cross. The beams are marked in
  the lower left-hand corner of each panel by an open ellipse. The
  direction of the CO outflow is marked by the straight line.
}
\end{figure}

\begin{figure}
\plotone{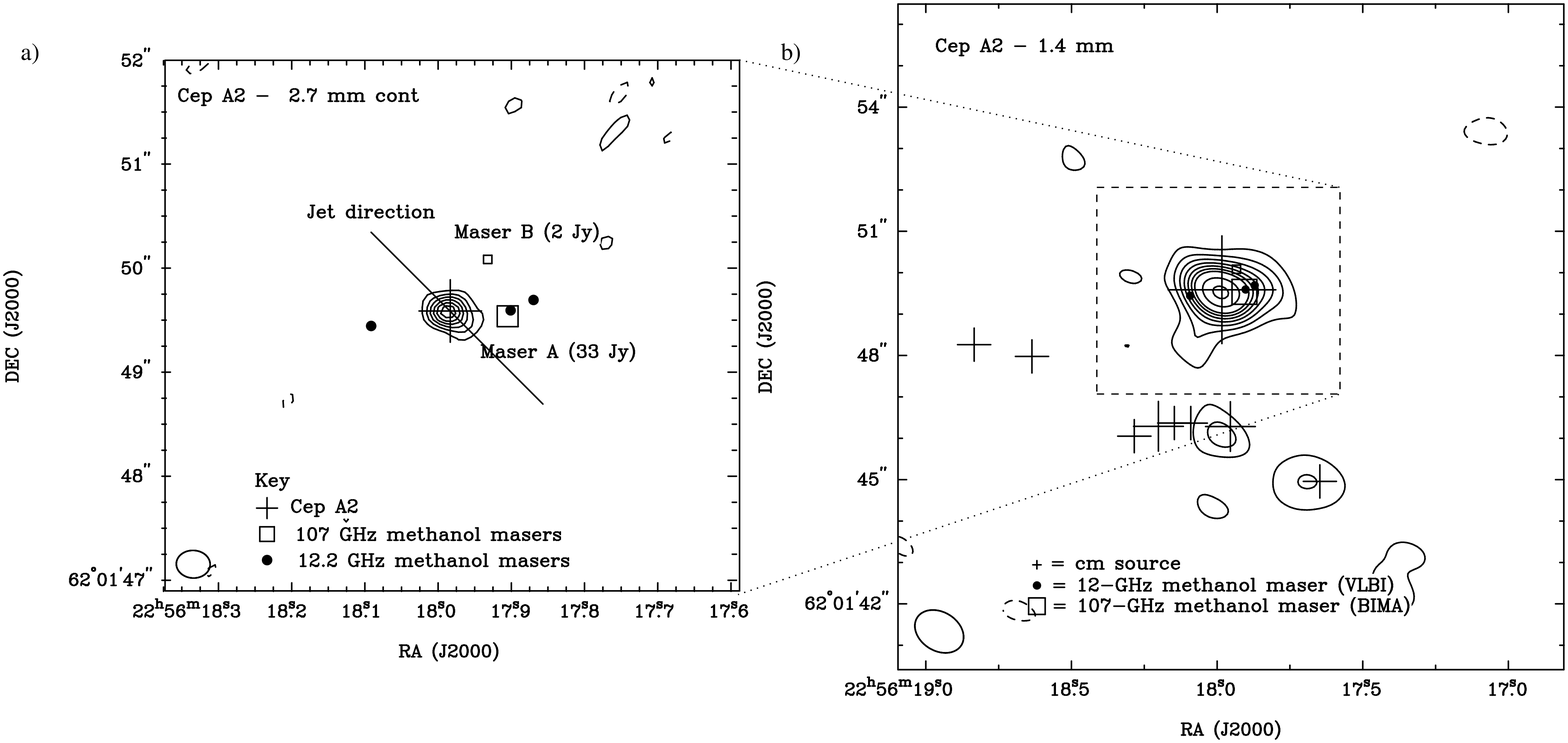}
\caption{Continuum emission towards Cep A at a) 2.7-mm and b) 1.4
  mm. Radio sources are marked by crosses, and methanol masers by
  filled circles (12 GHz) and open squares (107 GHz). The beams are
  marked in the lower left-hand corner of each panel. The direction of
  the radio jet is marked by the straight line. 
}
\end{figure}

\subsection{Cep A}
 
Cep A2 is the well-known radio-jet source in the Cep A region. The
2.7-mm continuum shown in Figure 
2a 
is slightly elongated along the direction of the jet
with FWHM dimensions of 0.45$\times$0.28 arcsec$^2$ corresponding to
315$\times$200 AU$^2$ at a distance of 700 pc. The emission at a
resolution of 0.3 arcsec remains dominated by the ionized jet although
images made using the more compact configurations reveal the presence
of dust emission from the envelope. At 1.4 mm (Fig. 
2b) 
the emission is
dominated by Cep A2 although there appears to be some extended
emission which encompasses the methanol masers to the west of Cep A2
with a further extension to the SE. This extended emission may
represent the remnant flattened core from which Cep A2 formed. A
Gaussian fit to the dust emission yields a slightly extended source
with deconvolved dimensions of 0.8$\times$0.5 arcsec$^2$
(560$\times$350 AU$^2$). Weak dust emission is also detected from two
of the Cep A3 group of sources (3b and 3c: Garay et al. 1996).

While the water masers mostly cluster around Cep A2 itself, the
methanol masers show no clear association with any particular
source. Our BIMA observations clearly resolve the two 107-GHz maser
spots of Mehringer et al.\ (1997) separate from Cep A2. The
stronger maser coincides with one of the 12.2 GHz methanol masers,
both spatially and spectrally.

\subsection{Other sources}

NGC7538-IRS1 is barely resolved at 2.7 mm: the `south spherical'
source of Gaume al. (1995) is discernible. The presence of this source
complicates the spectrum making it difficult to disentangle the
contributions from the dust and the ionized gas.

The strong 95-GHz methanol maser is detected towards GGD\,27-IRS and
resolved to be a single maser spot 9.2 arcsec north-north-east of the
jet source. This position places the maser along the axis of the jet
and it may arise in one of the jet shocks. The flux of GGD\,27 is
somewhat uncertain but is probably $\sim$50 mJy.

Not all of our mapped sources had strong compact emission at 2.7 mm.
NGC7538-IRS9 and G35.2$-$0.7N were only weakly detected at the highest
resolutions, and GL\,2591 was not detected at all to a 3-$\sigma$
sensitivity limit of 20 mJy in a 1-arcsec beam at 2.7 mm.

\section{Analysis: dust emission around massive YSOs}

Figure 3 plots the spectral energy distribution from radio through to
millimetre wavelengths for four of our sources. Radio measurements
were taken from the literature while the 2.7 and 1.4-mm points (filled
squares) are from our data. The radio spectral index has been used to
predict the contribution from the ionized wind at millimetre
wavelengths and is shown by the straight lines in Fig. 3. In each case
it is clear that dust emission has been detected even at 2.7 mm.

\begin{figure}[ht]
\plotone{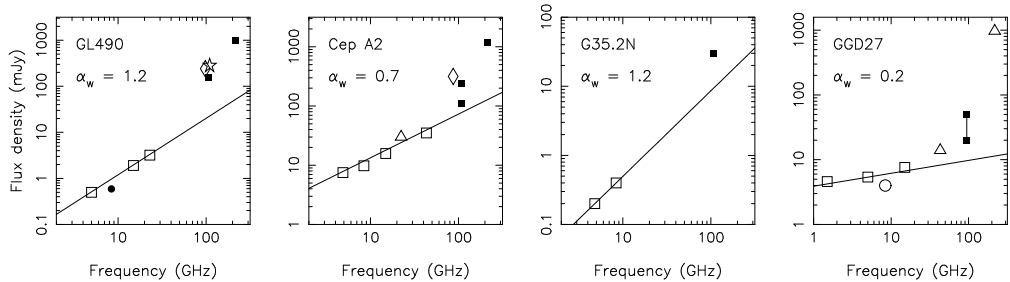}
\caption{Spectral energy distributions at wavelengths longer than 1.3
  mm. Our data are marked by filled squares; others are taken from the
  literature. The two points at 107 GHz for Cep A2 are for the A-array
  data alone (lower point) and A+B combined (upper point). The 2.7-mm
  flux for GGD27-IRS shows the range of values. The radio spectral
  index ($\alpha_{\rm w}$) is shown for each source.}
\end{figure}

Accurate mass estimates require detailed modelling, but here we
present simple estimates of the total mass assuming that the dust
emission is optically thin and is at a uniform temperature of 75 K
(see, e.g., Mundy \& Adelmann 1988). We take the dust emissivity to be
the same as that used by Looney et al.\ (2000) for low-mass sources.
Note that the mass estimates roughly scale linearly (but inversely)
with temperature. In all cases the mass estimate is of order 1--4
M$_\odot$ within a region $\sim$1 arcsec in diameter (see Table 1).

While it would be a nice result to be able to claim that we have
detected disks around these massive YSOs, in practice our resolution
and sensitivity are not high enough to determine whether the emission
comes from flattened disk-like structures or from a larger-scale
envelope. At 2.7 mm images of both GL\,490 and Cep A2 made with just
the very longest baselines of the A configuration show unresolved
sources ($<$0.2 arcsec FWHM) which are consistent with emission from
the ionized stellar wind. For GL\,490 the wind contribution of
$\sim$25 mJy allows us to place an upper limit on the mass within a
radius of 0.15$''$ (or 150 AU) of $\sim$0.5 M$_\odot$ (assuming the
same dust properties as Table 1). In order to separate the dust
emission on scales smaller than 0.2 arcsec, high-resolution
observations at 1.4-mm (or shorter wavelengths) are required. Even
then, discovery of a disk can not be claimed until a kinematic
signature is observed.  We are currently examining our line data but
so far have not detected a clear rotation signature.

\begin{table}
\centering
\caption{Estimates of circumstellar dust masses. Contributions to the
  flux density from the ionized wind have been subtracted. A dust
  temperature of 75 K has been assumed. }
\medskip
\begin{tabular}{lccccc}
\tableline
Source & \multicolumn{2}{c}{Flux (Jy)} & \multicolumn{2}{c}{Mass (M$_\odot$)} & Notes \\
       & 2.7 mm & 1.4 mm & 2.7 mm & 1.4 mm & \\
\tableline
GL\,490      &  ~~0.12 & 1.0 &  ~~2.5 & 2.6 & \\
GGD27-IRS1   &  ~~0.04 & --  &  ~~2.6 & --  & \\
G35.2$-$0.7N &  ~~0.02 & --  &  ~~1.8 & --  & \\
GL\,2591     & $<$0.02 & --  & $<$0.4 & --  & 1$''$ beam\\
Cep A2       &  ~~0.05 & 1.1 &  ~~0.5 & 1.4 & \\
NGC7538-IRS1 &  ~~0.87 & 2.6 &  ~~143 &  54 & \\
NGC7538-IRS9 &  ~~0.03 & --  &  ~~4.1 & --  & \\
\tableline
\tableline
\end{tabular}
\end{table}

\section{Conclusions: the search for disks goes on}

We have presented results from a search for dusty disks around a
sample of massive YSOs. We have found that while dust is clearly
present around these sources, the dust emission at 2.7 mm is too weak
to separate from the free-free emission from the ionized wind. Our
observations at 1.4-mm do not have sufficient resolution to probe the
geometry, but do show that several solar masses of dusty material
exist on scales of a 1000 to 2000 AU from the central source. The
exception is GL\,490 which we clearly resolve into an elongated
structure with dimensions 0.7$\times$0.5 arcsec$^2$, or 700$\times$500
AU$^2$. Although it is not possible for us to determine unambiguously
whether this structure represents a circumstellar disk or if we are
beginning to resolve a binary system, we note that the near-infrared
speckle imaging of Hoare et al.\ (1996) revealed no evidence for
multiplicity towards GL\,490.

Future observations at 2.7 mm will need significantly higher
sensitivity and resolution. A more promising approach is to employ
observations at 1.4 mm or shorter wavelengths where the wind
contribution will be negligible. In this respect the completion of
CARMA ({\tt http://www.mmarray.org/}) at its high-altitude site will
mark the beginning of a new era in the search for disks around massive
YSOs.

\subsection*{Acknowledgments}

BIMA is funded by NSF grant AST-9981289.

\end{document}